\documentclass[%
reprint,
superscriptaddress,
showpacs,
amsmath,amssymb,
aps,
prl,
longbibliography,
]{revtex4-1}
\usepackage{psfrag,graphicx,epsfig,color}
\usepackage{dcolumn}
\usepackage{bm}
\usepackage{natbib}
\usepackage{float}
\usepackage[usenames,dvipsnames,svgnames,table]{xcolor}
\usepackage{subfigure}
\usepackage{nicefrac}

\graphicspath{{figs/}{plots/}}

\def\re {{R_\lambda}}

\definecolor{mygreen}{rgb}{0,0.7,0.}

\begin{document}

\title{
Vorticity-strain rate dynamics
and the smallest scales of
turbulence
}

\author{Dhawal Buaria }
\email[]{dhawal.buaria@nyu.edu}
\affiliation{Tandon School of Engineering, New York University, New York, NY 11201, USA}
\affiliation{Max Planck Institute for Dynamics and Self-Organization, 37077 G\"ottingen, Germany}
\author{Alain Pumir}
\affiliation{Laboratoire de Physique, ENS de Lyon, Universit\'e de Lyon 1 and CNRS, 69007 Lyon, France}
\affiliation{Max Planck Institute for Dynamics and Self-Organization, 37077 G\"ottingen, Germany}

\date{\today}

\begin{abstract}

Building upon the intrinsic properties of Navier-Stokes dynamics, 
namely the prevalence of intense vortical structures and the
interrelationship between vorticity and strain rate,
we propose a simple framework to quantify the 
extreme events and the smallest scales of turbulence.
We demonstrate that our approach is
in excellent
agreement with the best available data from 
direct numerical simulations 
of isotropic turbulence,
with Taylor-scale Reynolds number
up to 1300.
We additionally highlight a shortcoming of 
prevailing intermittency models due
to their disconnection from observed
correlation between vorticity and strain.
Our work accentuates the importance of this
correlation as a crucial step in developing an accurate 
understanding of intermittency in turbulence.

\end{abstract}

\maketitle


A defining property of fluid turbulence
is the presence of a wide range of 
dynamically interacting scales -- bounded
from above by the largest scales,
which are of the order of flow
dimension, and from below by the smallest
scales, determined by the diffusive
action of molecular viscosity.
The largest scales transport the bulk
of the flow energy and momentum, 
whereas the smallest scales are responsible for
dissipating the flow energy into heat.
The net transfer of energy from large to 
small-scales (onto viscous dissipation) 
occurs via an energy cascade,
which renders the averaged 
energy dissipation-rate $\langle \epsilon \rangle$
to become independent of (kinematic) viscosity $\nu$
-- which is also termed dissipative anomaly.  
This phenomenology, first proposed
by Kolmogorov (1941) \cite{K41a, K41b} (K41 henceforth),
identifies the smallest scales in the flow as:
\begin{align}
\eta_{\rm K} = \left( \nu^3/ \langle \epsilon \rangle \right)^{1/4} 
\ \ ; \ \
\tau_{\rm K} = \left( \nu/ \langle \epsilon \rangle \right)^{1/2} 
\label{eq:kscale}
\end{align}
where $\eta_{\rm K}$ and $\tau_{\rm K}$ are the Kolmogorov
length and time-scale respectively.

While dissipative anomaly has been widely
confirmed \cite{sreeni84,sreeni98,pearson02,kaneda03},
the overall mean-field description 
of K41 has been invalidated \cite{Frisch95,Sreeni97}.
This is because the fluctuations of dissipation-rate,
and velocity gradients in general,
exhibit a high-degree of spatial and temporal
intermittency, with large
non-Gaussian excursions from its mean,
which become increasingly stronger 
with the Reynolds number
\cite{MS91, BPBY2019, Elsinga20}.
Such extreme events
play a crucial role in numerous
physical processes 
\cite{falkovich02, Sreeni04, vergassola07, hamlington12}
and are at the center of 
turbulence theories and models 
\cite{Frisch95, Sreeni97, Meneveau11}.
Simultaneously, it follows
that the smallest scales in the flow,
putatively corresponding to the extreme events, 
would be smaller than the Kolmogorov
scales defined by Eq.~\eqref{eq:kscale}
\cite{Paladin87, Sreeni88, Nelkin90, YS:05, BPBY2019}. 

Several phenomenological models have been proposed 
to describe intermittency, with reasonable success
in characterizing the statistics of velocity increments
for inertial-scales \cite{Frisch95, Sreeni97}. 
However, an accurate description of the
the smallest scales has 
remained elusive for two reasons.
The first limitation is the insufficiency of 
well-resolved data across a wide range of Reynolds numbers, 
since capturing extreme events requires a 
very stringent small-scale resolution 
\cite{Donzis:08, PK+18, BPBY2019}.
The second limitation is that of phenomenological
models, which typically appeal to adjustable parameters
without a clear connection
to the dynamics of the Navier-Stokes equations.
For instance, it 
is well-known that extreme gradients
are structurally arranged
in tube-like vortices 
\cite{Douady:91, Jimenez93, Ishihara09,BPBY2019}; 
however,  prevailing
intermittency theories neither predict
this feature nor take it into account in a 
precise manner.

In this Letter, overcoming
the aforementioned limitations, we propose a
simple framework to characterize the smallest
scales of turbulence based on vortical flow structures
while also directly connecting to the
underlying Navier-Stokes dynamics.
Our predictions are validated with data from 
state-of-the-art direct numerical 
simulations (DNS) of the incompressible 
Navier-Stokes equations,
demonstrating excellent agreement.
We additionally show that predictions
from prior intermittency models are recovered 
as a limiting case of our framework,
and discuss the root of this discrepancy.

The DNS data utilized here correspond to
the canonical setup of
forced stationary isotropic turbulence 
in a periodic domain \cite{Ishihara09},
enabling the use of highly accurate 
Fourier pseudo-spectral methods \cite{Rogallo}.
The key novelty of our data is that we have
simultaneously achieved both very 
high Reynolds number and the prescribed
small-scale resolution to accurately
resolve the extreme events \cite{PK+18, BPBY2019}.
The data correspond to the same Taylor-scale 
Reynolds number $\re$ range of $140-1300$
as attained in recent studies
\citep{BS2020, BBP2020, BPB2020, BP2021, BPB2021, buaria.cpc}.
However, the run at $\re=1300$ is extended
to a grid of $18432^3$ points (see \cite{YR2020}) --
the largest DNS run to date -- presenting
a substantial improvement on any previous
work investigating the smallest scales
\cite{schum07sub, Donzis:08, BPBY2019}. 
The resolution is
$k_{\rm {max}} \eta_{\rm K} \approx 6$
for $\re \le 650$, and
$k_{\rm {max}} \eta_{\rm K} \approx 4.5$
for $\re=1300$, where
$k_{\rm {max}} = \sqrt{2}N/3$, is the maximum
resolved wavenumber on a $N^3$ grid.
Convergence with respect to resolution and statistical 
sampling has been thoroughly established in 
previous works
\citep{PK+18, BPBY2019, YR2020}.

\begin{figure}
\begin{center}
\hspace{-0.2cm}
\includegraphics[width=0.24\textwidth]{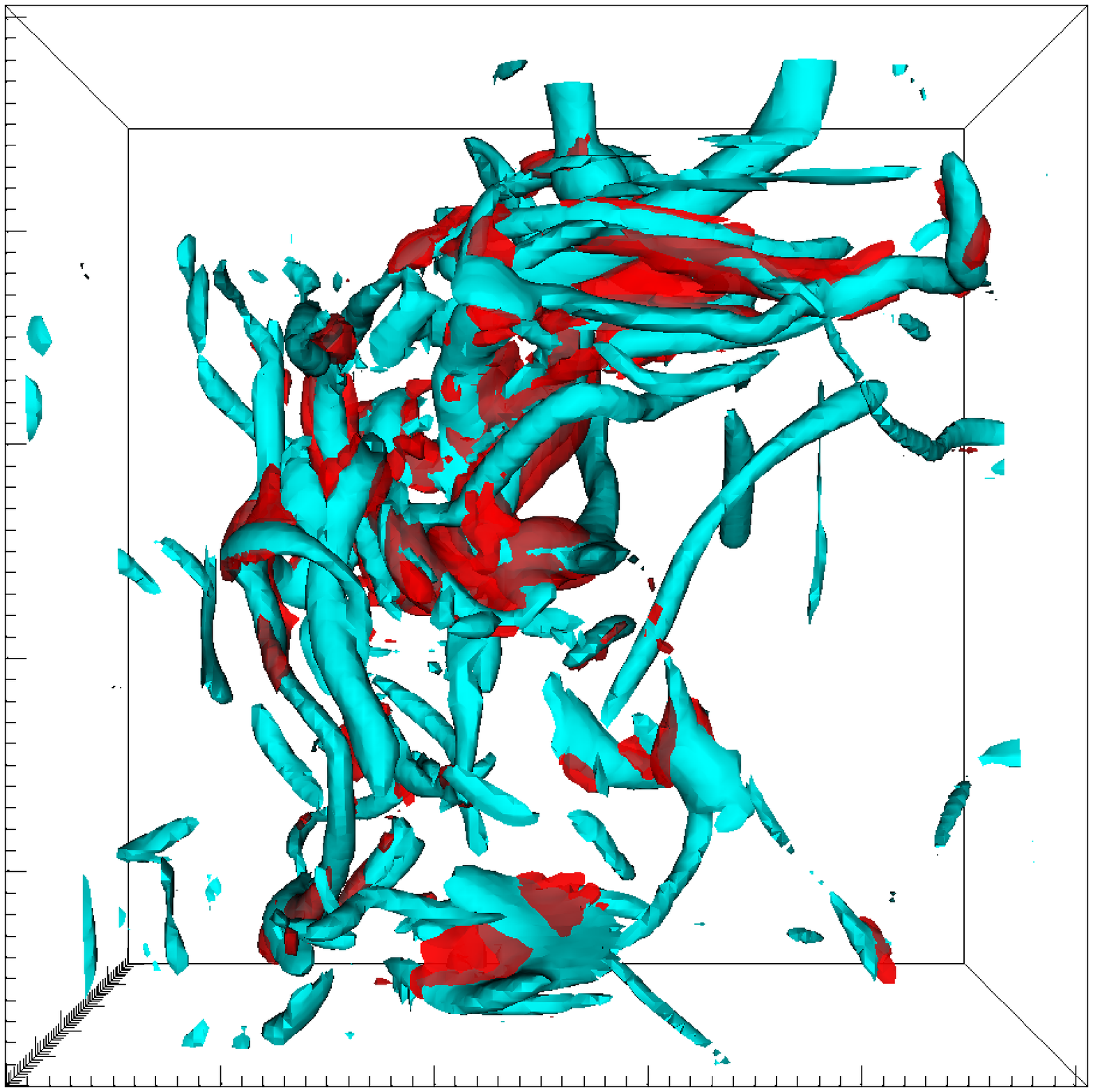} 
\hspace{-0.2cm}
\includegraphics[width=0.24\textwidth]{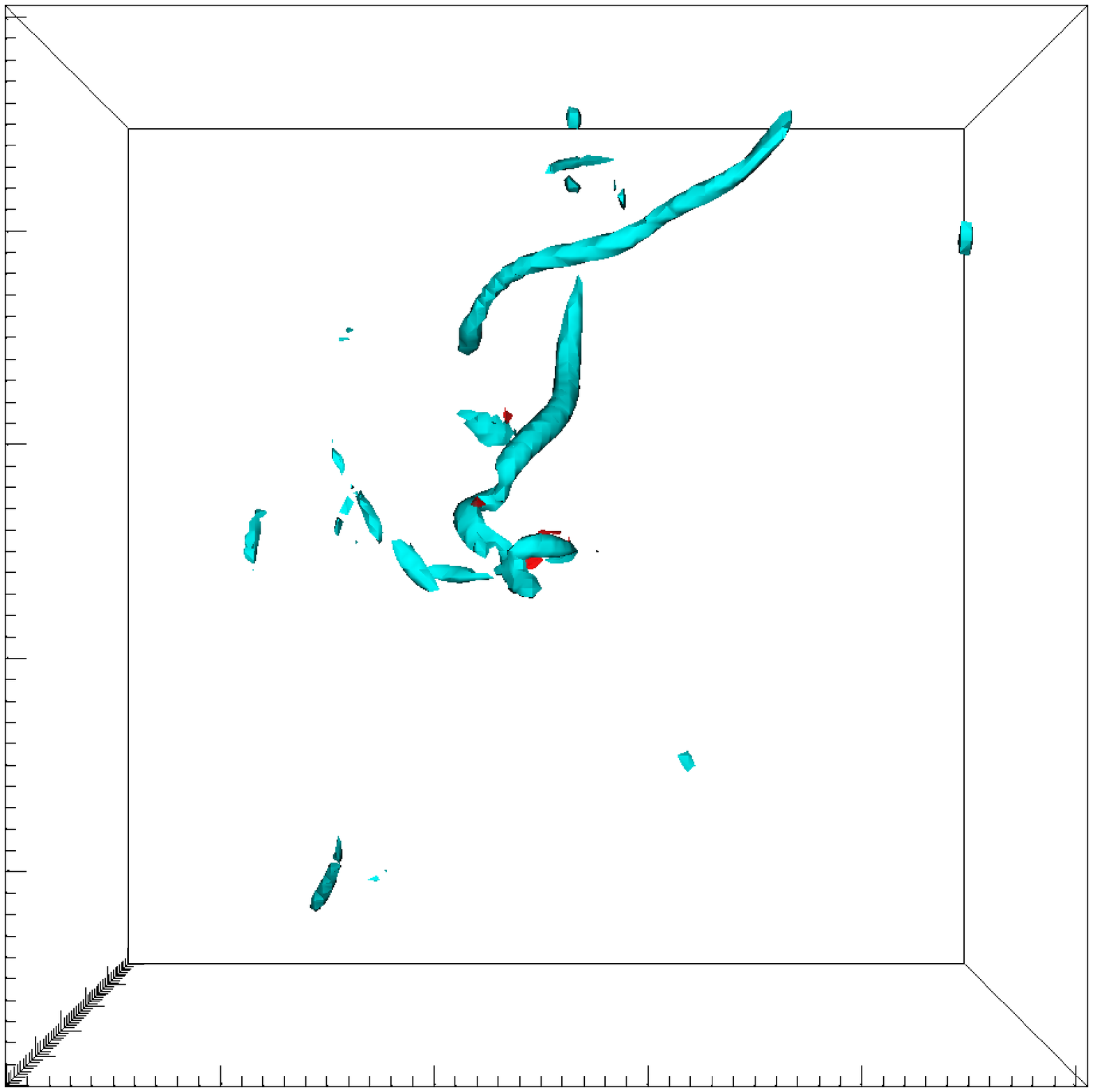} 
\caption{
3D-contour surfaces
of $\Omega \tau_{\rm K}^2$ (cyan) and
$\Sigma \tau_{\rm K}^2$ (red) at $\re=1300$.
The contour thresholds are $500$ (left) and
$4000$ (right) respectively, and the domain
size is $(100 \ \eta_{\rm K})^3$.
}
\label{fig:vis}
\end{center}
\end{figure}

We first identify the intrinsic features of
Navier-Stokes dynamics that are essential 
to characterize the small-scales.
From the velocity gradient tensor
$A_{ij} =\partial u_i/\partial x_j$,
two important descriptors of small-scale motions
can be identified, viz., the
strain-rate tensor $S_{ij} = (A_{ij} + A_{ji})/2$,
and the vorticity vector $\omega_i = \epsilon_{ijk} A_{jk}$
($\epsilon_{ijk}$ being the Levi-Civita symbol). 
We utilize their square-norms:
\begin{align}
\Sigma = 2S_{ij}S_{ij} \ , \ \ \ \Omega = \omega_i \omega_i \ , 
\end{align}
the former being the dissipation-rate without the
viscosity, i.e., $\Sigma = \epsilon/\nu$, and $\Omega$ 
being the enstrophy. From statistical homogeneity and 
the definition of $\tau_{\rm K}$, it follows 
$\langle \Omega \rangle = \langle \Sigma \rangle = 1/\tau_{\rm K}^2 $. 
It is well-known that the interaction of strain and vorticity 
plays a direct role in generating extreme gradients in the flow
and hence the smallest scales \cite{Jimenez93, Tsi2009}. 
Understanding the salient properties of this interaction 
constitutes the first step of our analysis.

Figure~\ref{fig:vis} shows the structure of extreme
events at the highest $\re$ ($=1300$), via visualization
of isosurfaces of strain and vorticity.  
Figure~\ref{fig:vis}a corresponds to a moderately large threshold
and illustrates the well-known picture of intense gradients
corresponding to vortical filaments, surrounded by
sheet-like regions of intense strain  
\cite{Jimenez93, moisy04, Ishihara09, BPBY2019}.
In Fig.~\ref{fig:vis}b, a substantially larger threshold
is chosen, and remarkably  
vortical filaments
still prevail.   

\begin{figure}
\begin{center}
\includegraphics[width=0.45\textwidth]{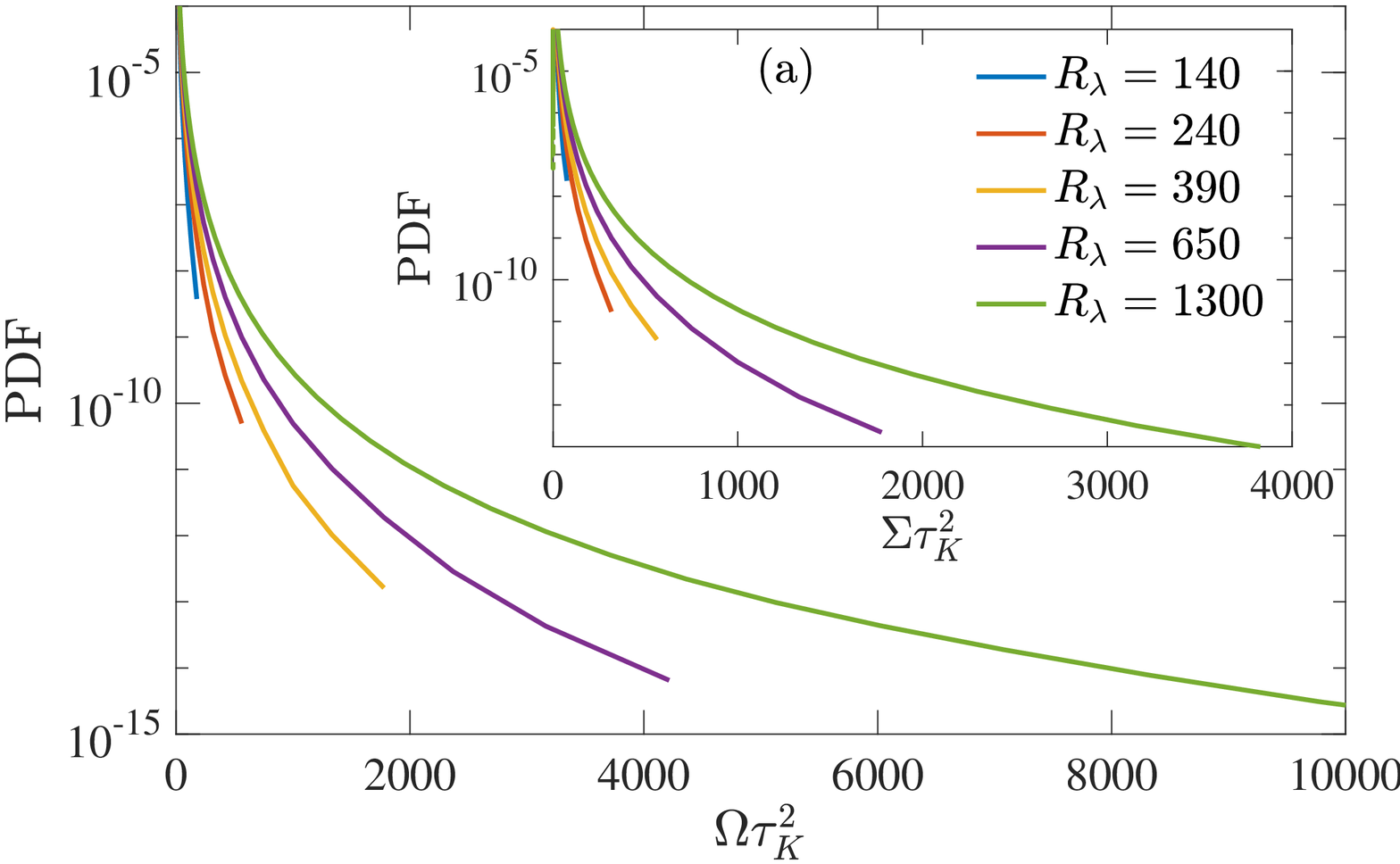} \\ 
\vspace{0.2cm}
\includegraphics[width=0.45\textwidth]{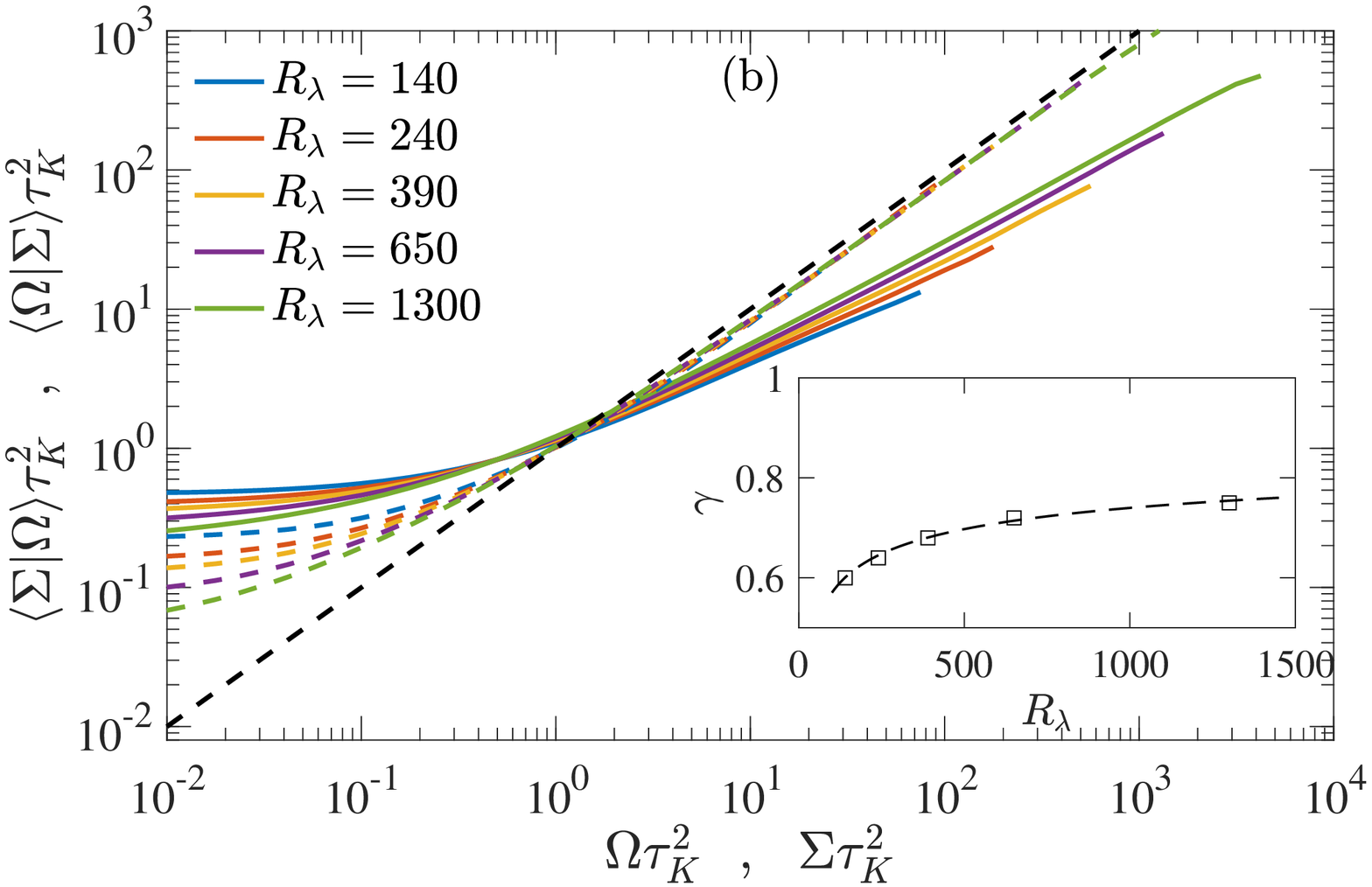}
\caption{
(a) Probability density functions (PDFs) of
$\Omega \tau_{\rm K}^2$ and $\Sigma \tau_{\rm K}^2$ (inset)
for various $\re$.
(b) Conditional expectations 
$\langle \Sigma | \Omega \rangle$ (solid lines)
and  
$\langle \Omega | \Sigma \rangle$ (dashed lines)
for various $\re$. The black dashed line corresponds 
to a slope of 1.
Inset shows $\gamma$ as a function of $\re$, 
for a power-law $\langle \Sigma | \Omega \rangle \sim \Omega^\gamma$ 
applied in the region $\Omega\tau_{\rm K}^2 \gtrsim 1$. 
}
\label{fig:pdf1}
\end{center}
\end{figure}

Although $\Omega$ and $\Sigma$ have the same mean,
it is well-known that $\Omega$ is more intermittent
\cite{kerr85, Ishihara09, BPBY2019}
-- likely due to the disparate role of vortex stretching
in amplifying vorticity and simultaneously
depleting strain \cite{Tsi2009, BPB2021}.
This is reflected in 
their probability density functions (PDFs) 
in Fig.~\ref{fig:pdf1}a,
which firmly establishes 
that this difference
is not a low-$\re$ effect as previously 
believed \cite{he98, nelkin99, Donzis:08}.
The local interrelationship between strain and vorticity
can be better understood by considering their
mutual conditional expectations as
shown in Fig.~\ref{fig:pdf1}b (and also have been
studied previously in various contexts 
\cite{Donzis:08, BPBY2019, BP2021}). 
The main observation is that while large 
$\Sigma$ is accompanied by
proportionately large $\Omega$,
i.e., $\langle \Omega |\Sigma \rangle \sim \Sigma$,
the converse is not true. Instead,
the strain in regions of intense vorticity
is considerably weaker, and empirically
described by the power-law:
\begin{align}
\langle \Sigma | \Omega \rangle \tau_{\rm K}^2  \sim (\Omega \tau_{\rm K}^2)^\gamma  \ \ , 
\  0< \gamma < 1
\label{eq:gamma}
\end{align}
where the exponent $\gamma$ 
slowly increases with $\re$
(see inset of Fig.~\ref{fig:pdf1}b). 
Notably, existing intermittency models
neither predict this, nor take it into
account when characterizing the smallest scales.

With the knowledge of vortical flow structures 
and the asymmetry between the behavior
of strain and vorticity (reflected in the exponent
$\gamma$), we now formulate the framework to 
quantify the smallest scales in the flow. 
From a physical standpoint, 
the smallest length-scale in the flow
corresponds to the smallest 
dimension of vortical structures, 
as obtained from a balance  
between viscosity and some effective strain 
$S_{\rm e} \simeq \Sigma_{\rm e}^{1/2}$
acting on the
particular structure \cite{Burgers48}:
\begin{align}
\eta = (\nu^2/\Sigma_{\rm e})^{1/4}
\label{eq:Burg}
\end{align}
which can be rewritten as
\begin{align}
\eta /\eta_{\rm K} = (\Sigma_{\rm e} \tau_{\rm K}^2)^{-1/4}
\end{align}
The classical Kolmogorov result in Eq.~\eqref{eq:kscale}
is obtained for $\Sigma_{\rm e}$ corresponding to 
the mean-field, i.e., 
$\Sigma_{\rm e} = \langle \epsilon \rangle /\nu = 1/\tau_{\rm K}^2$.
Instead, the observation in Fig.~\ref{fig:pdf1}
suggests utilizing the conditional
relation in Eq.~\eqref{eq:gamma}, leading to
\begin{align}
\eta /\eta_{\rm K} = (\Omega \tau_{\rm K}^2)^{-\gamma/4}
\end{align}
Introducing the length-scale $\eta_{\rm {ext}}$ 
as the size associated with vortex structures 
corresponding to $\Omega_{\rm {max}}$, which in turn
corresponds to the smallest time-scale $\tau_{\rm {ext}}$,
i.e., $\Omega_{\rm {max}} \sim \tau_{\rm {ext}}^{-2}$,
leads to
\begin{align}
\eta_{\rm {ext}} /\eta_{\rm K} = (\tau_{\rm {ext}}/\tau_{\rm K})^{\gamma/2}
\label{eq:et}
\end{align}
Keeping in mind the growth of PDF tails 
with $\re$ (when normalized by $\tau_{\rm K}$), 
we can write:
\begin{align}
\eta_{\rm {ext}} = \eta_{\rm K} \times \re^{-\alpha} \ , \ \ \ 
\tau_{\rm {ext}} = \tau_{\rm K} \times \re^{-\beta}  \ ,
\end{align}
where $\alpha, \beta >0$ are to be determined.
Substituting these 
in Eq.~\eqref{eq:et} leads to
\begin{align}
2 \alpha  = \gamma \beta \ ,
\label{eq:abc1}
\end{align}
giving first direct relation between 
$\alpha$ and $\beta$.

We now recall
that velocity gradients
in the flow simply correspond to velocity 
increments across the smallest length-scale.
Hence, the strongest gradient
simply corresponds to the largest
velocity increment, say $\delta u_{\rm {max}}$ 
over $\eta_{\rm {ext}}$: 
\begin{align}
1/ \tau_{\rm {ext}} \sim  \delta u_{\rm {max}} / \eta_{\rm {ext}} \ .
\label{eq:dumax}
\end{align}
Based on earlier works 
\cite{Paladin87, Jimenez93, BPBY2019}
(see also \cite{supp}): 
\begin{align}
\delta u_{\rm {max}} \sim u^\prime \ ,
\end{align}
where $u^\prime$ is the r.m.s. of velocity;
which upon substitution in Eq.~\eqref{eq:dumax} gives 
\begin{align}
\beta = \alpha + 1/2 \ ,
\label{eq:abc2}
\end{align}
Here, we have used the standard estimate
$u^\prime /u_{\rm K} \sim \re^{1/2}$, where
$u_{\rm K} = \eta_{\rm K}/\tau_{\rm K}$. 
Finally, solving Eqs.~\eqref{eq:abc1} 
and \eqref{eq:abc2} allows us to
obtain $\alpha$ and $\beta$ in terms of $\gamma$:
\begin{align}
\beta = \frac{1}{2-\gamma} \ , \ \  
\alpha = \frac{\gamma}{2(2-\gamma)} \ .
\label{eq:abc3}
\end{align}

\begin{figure}[h]
\begin{center}
\includegraphics[width=0.44\textwidth]{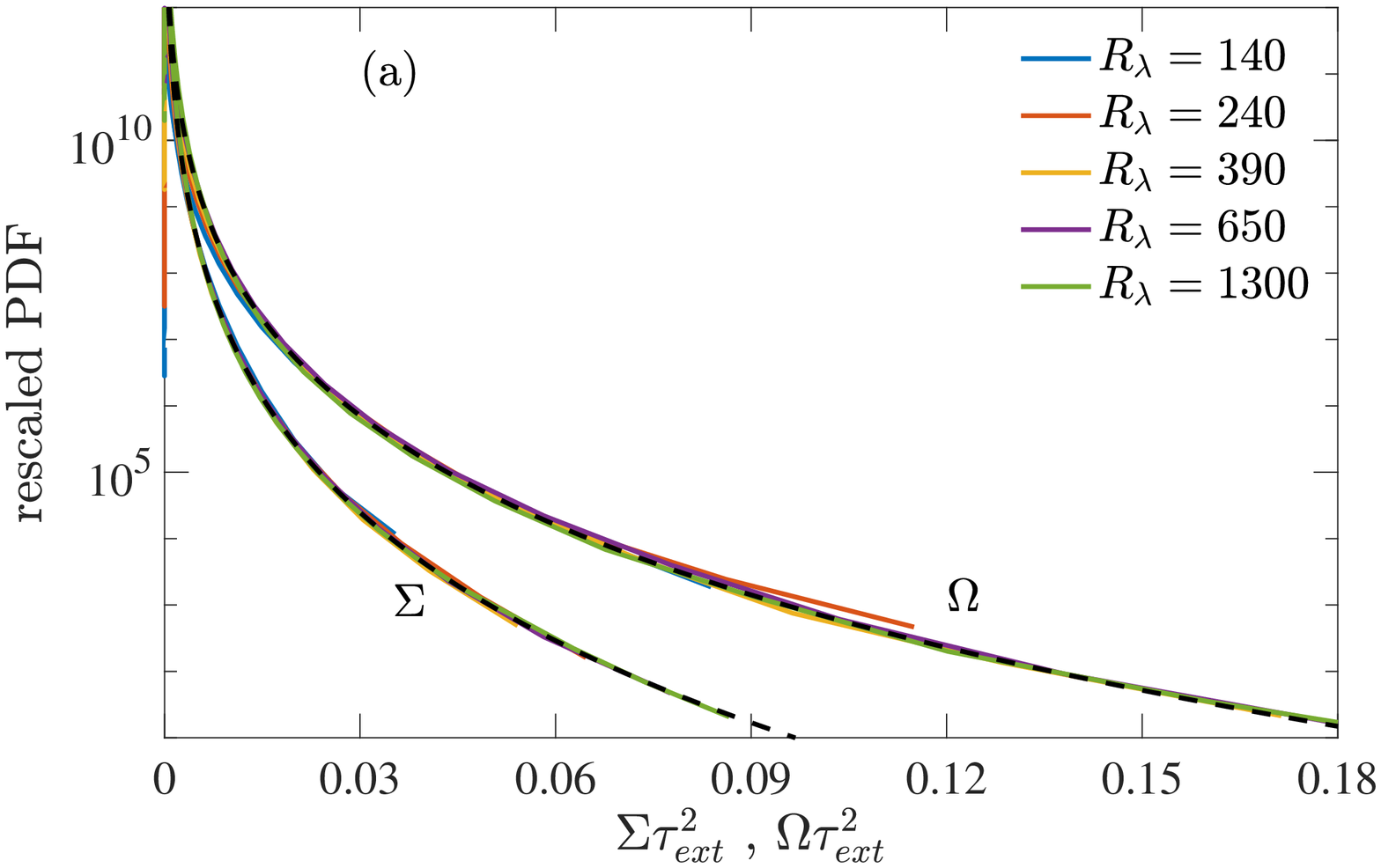} \\
\vspace{0.6cm}
\includegraphics[width=0.44\textwidth]{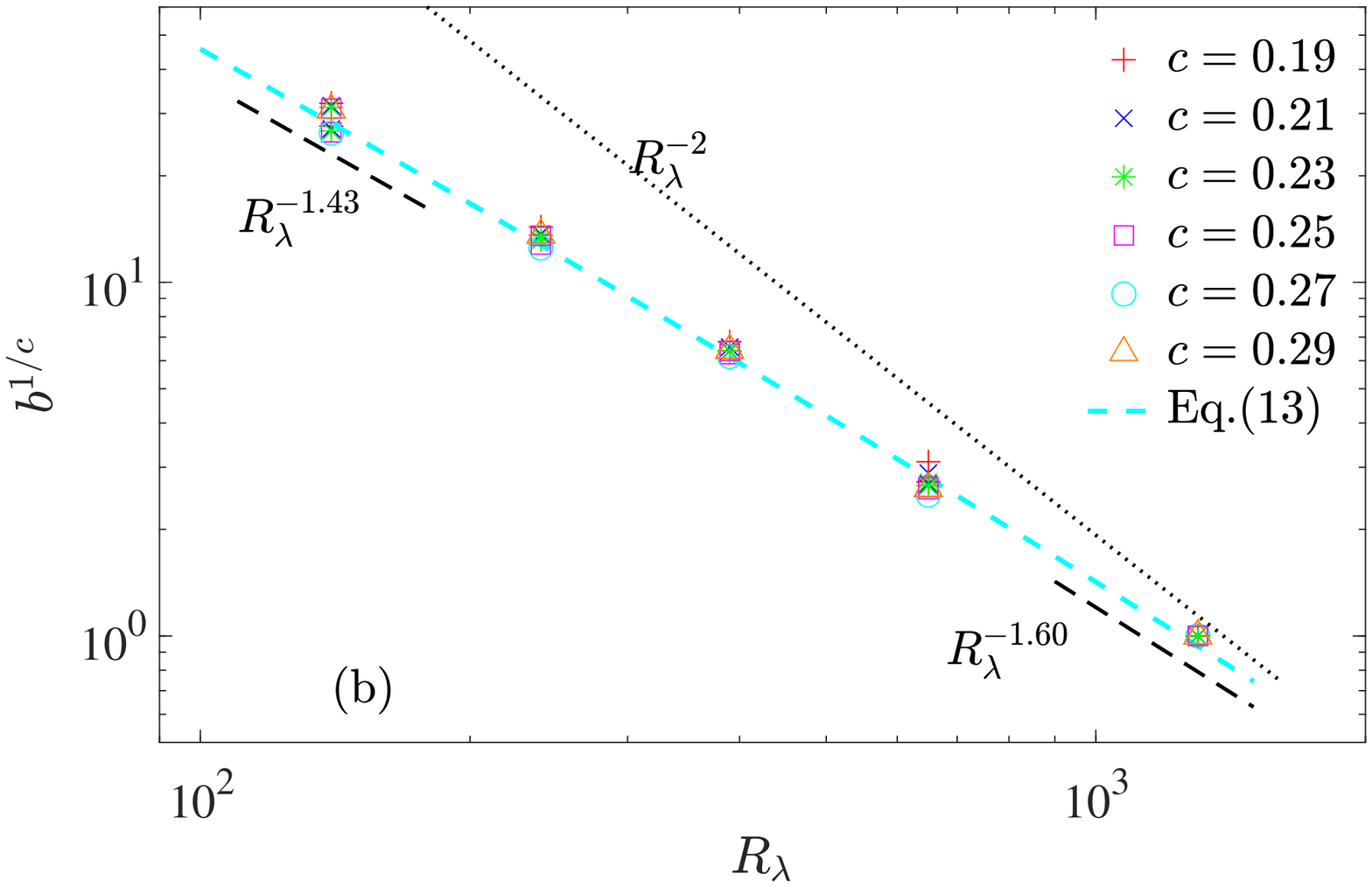}
\caption{
(a) Rescaled PDFs of $\Omega$ and $\Sigma$,
normalized by $\tau_{\rm {ext}}^2$ corresponding
to smallest time-scale, at various $\re$.
The black dashed line corresponds
to stretched exponential fit.
(b) Plot of $b^{1/c}$ vs. $\re$
corresponding to stretched exponential fits to
PDFs of $\Omega$ and $\Sigma$.
For clarity, we have rescaled the curves,
so all data point superpose at $\re=1300$.
The dashed (cyan) line corresponds
to the prediction for $\beta$ in 
Eq.~(13), taking into account the variation of $\gamma$ 
shown in inset of Fig.~\ref{fig:pdf1}b \cite{supp},
whereas the dotted line
corresponds to the prediction
$\beta=1$ in Eq.~(19).
}
\label{fig:pdf2}
\end{center}
\end{figure}

To first validate the result for $\beta$,
we return to the PDFs shown 
in Fig.~\ref{fig:pdf1}a, with
the expectation that rescaling them
with $\tau_{\rm {ext}}$ should 
collapse the tails.
Figure~\ref{fig:pdf2}a shows this
result, with $\beta$ (and $\tau_{\rm {ext}}$) defined based
on $\gamma$ obtained in Fig.~\ref{fig:pdf1}b
-- demonstrating excellent
agreement. 
It should be noted that a similar collapse was 
also obtained in \cite{BPBY2019}
at lower $\re$ and for a fixed value of $\beta=0.775$.
However, the current data at significantly higher $\re$
negate a fixed value of $\beta$.
Hence, the $\re$-dependence of $\beta$ (arising from 
$\gamma$) is a crucial
ingredient of the current approach and
imperative for obtaining an accurate description.
This expectation and the quantitative 
variation of $\beta$ are also consistent 
with recent results of \cite{Elsinga20}, 
which characterize the scaling
of extreme dissipation events based on underlying
shear-layer structures. 

While the arguments leading
to Eq.~\eqref{eq:abc3}  
utilize the physical picture of
intense vorticity tubes, 
the collapse in 
Fig.~\ref{fig:pdf2}a remarkably indicates
that both extreme vorticity
and strain scale with
$\tau_{ext}$ -- though these
extrema arise from different spatial locations. 
This suggests that the amplification
of vorticity and strain 
occurs simultaneously with the same
time-scale, albeit non-locally
\footnote{
note, vorticity and strain are coupled
non-locally via the Biot-Savart relation
\cite{BPB2020, BP2021}},  
inducing the asymmetry in
their local correlation
as observed in Fig.~\ref{fig:pdf1}b.
Thus, the exponent 
$\gamma < 1$, which captures
this asymmetry, also captures the 
non-locality of vorticity-strain 
interaction. In fact, as demonstrated later,
this is also the key reason
for our framework's success over prior
intermittency models.

The collapse in Fig.~\ref{fig:pdf2}a can be additionally 
verified by noting that the tails of
PDFs of $\Sigma$ and $\Omega$
(and velocity gradients in general) are well 
fitted by stretched-exponential functions
\cite{MS91, benzi91, KSS92, zeff:2003, Donzis:08, BPBY2019}:
\begin{align}
f_X(x) \approx a \exp(-b \ x^c) \ ,
\label{eq:fx}
\end{align}
where $x = \Omega \tau_{\rm K}^2$ or $\Sigma \tau_{\rm K}^2$.
Applying a change of variable 
$x_{\rm e} = x (\tau_{\rm K}/\tau_{\rm {ext}})^2$ 
will lead to the transformation 
$b \rightarrow b' = b \times \re^{2 \beta c}$.
A necessary condition to collapse the tails would 
imply that $b'$ is independent of $\re$,
leading to the expectation that 
$b^{1/c} \sim \re^{-2\beta}$
(for any given value of $c$).


Figure~\ref{fig:pdf2}b shows the plot of $b^{1/c}$
as a function of $\re$,
for various $c$ values
(comprehensive details about the fitting 
procedure, and the chosen range of $c$ are 
discussed in the Supplementary~\cite{supp}).
We compare the slope of the data points
for $b^{1/c}$ with the result for $\beta$ in 
Eq.~\eqref{eq:abc3} by utilizing the $\gamma$ 
obtained earlier from Fig.~\ref{fig:pdf1}b
(note $\gamma$ varies from
$0.60-0.75$ for $\re=140-1300$) -- once again,
demonstrating excellent agreement. 
It is worth iterating that the collapse
obtained in Fig.~\ref{fig:pdf2}a 
does not depend on the curve fitting procedure.
Nevertheless,
this fitting procedure
independently reaffirms the 
robustness of our result and rules out any ambiguity.

While the result for $\tau_{\rm {ext}}$ (and $\beta$) 
was readily verified using
the PDF tails, verifying $\eta_{\rm {ext}}$ (and $\alpha$)
presents an inherent difficulty.
A simple approach would be to evaluate the PDF 
of the coarse-grained gradient $\delta u_r/r$,
where $\delta u_r$ is the velocity increment over some scale $r$,
and successively make $r$ smaller until the PDF of $\delta u_r/r$
collapses to the PDF of velocity gradient for 
$r \le \eta_{\rm {ext}}$. 
However, DNS data only provides 
discrete values of $r$ (in integer multiples of the grid 
spacing $\Delta x$), making it impractical
to precisely identify the exact $r/\eta_{\rm {ext}}$ 
without invoking some interpolation or approximate analysis.
Instead, we devise a simple alternative by 
characterizing the deviations of $\delta u_r/r$ from the actual
gradient. Note, the velocity increment can be longitudinal
or transverse, i.e., corresponding to velocity component
parallel or perpendicular (respectively) to the separation
vector, but it will be evident that
this choice is immaterial.

\begin{figure}
\begin{center}
\includegraphics[width=0.45\textwidth]{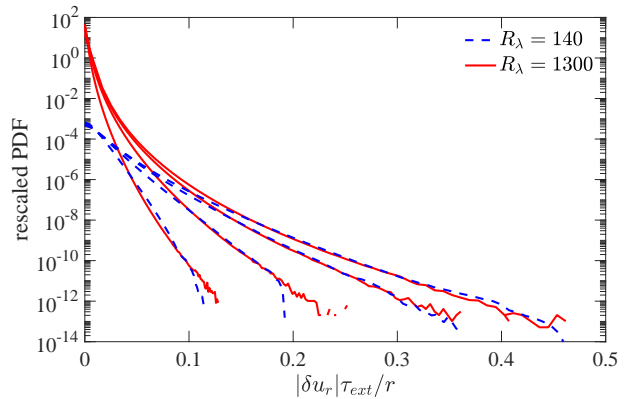}
\caption{Rescaled PDFs of the (transverse) 
velocity increments, $\delta u_r$, 
non-dimensionalized by $\tau_{\rm {ext}}/r$.
Solid red lines are for $\re=1300$,
showing $r/\eta_{\rm K}=1,2,4,8$,
and dashed blue lines are for $\re=140$,
showing 
$r/\eta_{\rm K}=2,4,8,16$,
corresponding to the ratio
of $\eta_{\rm K}/\eta_{\rm {ext}}$ for these two $\re$.
Curves for increasing $r/\eta_{\rm K}$ shift monotonically 
from right to left. Although not shown, 
the curves corresponding to the longitudinal 
increments exhibit similar behavior.
}
\label{fig:PDF_dv_ov_ueta2}
\end{center}
\end{figure}

From the Taylor-series expansion of $\delta u_r$, it 
follows:
\begin{align}
\frac{\delta u_r}{r} = 
\frac{\partial u}{\partial x} +  
\frac{\partial^2 u}{\partial x^2} \frac{r}{2!} +
\frac{\partial^3 u}{\partial x^3} \frac{r^2}{3!} + ...
\end{align}
For $r \le \eta_{\rm {ext}}$, the r.h.s. 
converges to $\partial u/\partial x$,
whereas for $r > \eta_{\rm {ext}}$
deviations are expected
due to the higher-order corrections.
For the most extreme gradients, 
we can nominally write:
$\partial^n u /\partial x^n \simeq c_n u^\prime/\eta_{\rm {ext}}^n$,
where $c_n$ are independent of $\re$.
Together with 
$u^\prime \sim \eta_{\rm {ext}}/\tau_{\rm {ext}}$,
this gives:
\begin{align}
\frac{\delta u_r \tau_{\rm {ext}}}{r} \simeq 
c_1 + 
\frac{c_2}{2!} \left( \frac{r}{\eta_{\rm {ext}}} \right) +
\frac{c_3}{3!} \left( \frac{r}{\eta_{\rm {ext}}} \right)^2 + ... \ ,
\end{align}
leading to expectation that the tails of PDFs of
$\delta u_r \tau_{\rm {ext}}/r$ can be collapsed
at different $\re$ by 
matching the $r/\eta_{\rm {ext}}$, thus providing a simpler test
with DNS data. 
We notice that 
$\eta_{\rm K}/\eta_{\rm {ext}}$
at $\re=140$ is approximately $2$ times 
than that at $\re=1300$.  
Fig.~\ref{fig:PDF_dv_ov_ueta2}
shows the rescaled PDF of 
$\delta u_r \tau_{\rm {ext}}/r$,
for $r/\eta_{\rm K}=1,2,4,8$ at $\re=1300$,
and $r/\eta_{\rm K}=2,4,8,16$ at $\re=140$,
showing a remarkably good collapse of the tails,
providing a strong confirmation of our approach.
Similarly, for $\re=140$ and $\re=650$,
the ratio of their $\eta_{\rm K}/\eta_{\rm {ext}}$ is approximately
$1.5$, and a similar collapse is obtained 
(see \cite{supp}).

The framework
developed in this work differs 
from previous phenomenological models, 
which ignore important features of the
Navier-Stokes dynamics.
In this regard,
the commonly utilized notion
is that the viscous cutoff scale
is defined by the phenomenological 
criteria of local Reynolds number 
being unity
\cite{Paladin87, Nelkin90, Frisch95, YS:05, schum07sub}: 
\begin{align}
\delta u_r  \ r/\nu \simeq 1 \ . 
\label{eq:re1}
\end{align} 
This is essentially an ad-hoc extension of 
the K41 phenomenology, since
$u_{\rm K} \eta_{\rm K}/\nu = 1$.
The velocity increment $\delta u_r$ 
is assumed to be H\"older continuous,
akin to multifractility \cite{Frisch95}:
\begin{align}
\delta u_r /u^\prime \sim  (r/L)^h  \ , 
\label{eq:holder}
\end{align} 
where $L$ is the large-eddy length
and $h$ is the local H\"older exponent.
It trivially follows that the smallest scales 
correspond to the minimum H\"older exponent $h_{min}$.  
Since $\delta u_r \sim u^\prime$
for the smallest scales, 
or equivalently $h_{min}=0$
\cite{Paladin87, Nelkin90, Frisch95},
it can be shown that \cite{supp}:
\begin{align}
\beta = 2 \alpha \ , \ \  
\beta = \alpha + \frac{1}{2}  \ ,
\label{eq:mf}
\end{align}
giving $\beta=1$ and $\alpha=1/2$,
in line with previous predictions 
\cite{Paladin87, Sreeni88, Nelkin90, YS:05}.

It can be readily seen that the result in 
Eq.~\eqref{eq:mf} differs from our results in
Eqs.~\eqref{eq:abc1} and \eqref{eq:abc2} 
only by the factor $\gamma$, 
both being the same if
$\gamma = 1$, i.e. when 
strain and vorticity are locally commensurate. 
While the numerical results
clearly demonstrate that $\gamma < 1$,
the weak increase in $\gamma$ with $\re$ 
(inset of Fig.~\ref{fig:pdf1}b)
is suggestive of a slow approach 
to $\gamma=1$ when $\re\to\infty$.
However, a nominal extrapolation of the data
in Fig.~\ref{fig:pdf1}b suggests that this limit
will be reached at extremely large $\re$,
beyond what can be achieved experimentally
or numerically \cite{supp}.
In fact, this is in line with 
previous and recent results
which independently 
reaffirm the shortcomings of the multifractal
model \cite{Jimenez93, BS2020, Elsinga20}.

Since the result in Eq.~\eqref{eq:holder}
(for $h=0$, giving 
$\delta u_r \sim u^\prime$) 
is consistent across all descriptions, the 
noted discrepancy 
arises from the criterion in 
Eq.~\eqref{eq:re1}.
For vortex tubes, the smallest scale
as set by Eq.~\eqref{eq:Burg} 
is qualitatively similar to the 
criteria in Eq.~\eqref{eq:re1}. However, it
does not provide any constraint
on the circulation of the vortex, $\Gamma$,
implying that the local Reynolds number 
defined as  $R_\Gamma = \Gamma/\nu$ is
not necessarily unity. 
Instead, our results indicate 
$R_\Gamma \simeq \re^{1 - \beta}$, in qualitative agreement
with the results of \cite{Jimenez93}.
Note, for $\re\to\infty$,  
the expectation  $\gamma, \beta \to 1$ 
implies that $R_\Gamma \to$ constant. 
Thus, the crucial misstep in prevailing intermittency models
appears to be its inability in distinguishing
longitudinal and transverse components
and utilizing their local correlation.
In fact, previous and recent results have shown that this
shortcoming also extends to inertial range,
where longitudinal
and transverse structure functions exhibit different
scaling exponents, contrary to the 
prediction from the multifractal model 
\cite{dhruva97, shen2002, gotoh02, grauer2012}.

In conclusion, we have developed a simple framework
to characterize the smallest scales of turbulence, 
which utilizes the underlying 
asymmetry between strain and vorticity dynamics
of Navier-Stokes equations.
We have demonstrated excellent agreement of DNS data with
predictions, and shown that our parametrization
reduces to predictions from existing
intermittency models when the symmetry between strain
and vorticity is restored,
albeit at extremely large $\re$,
which are unattainable on Earth.
In this regard, understanding the asymmetry
between vorticity and strain appears
to be a crucial component to understand intermittency
in turbulence for all practical situations
of interest, suggestive of a new avenue of investigation.
It would also be pertinent to extend 
the current framework to turbulent mixing of scalars,
where recent results have suggested that the smallest
scales in the scalar field also deviate from 
classical predictions \cite{BCSY2021a,BCSY2021b}.

\begin{acknowledgements}

\paragraph{Acknowledgments:}
We gratefully acknowledge the Gauss Centre for Supercomputing 
e.V. for providing computing time on the supercomputers 
JUQUEEN and JUWELS at J\"ulich Supercomputing Centre (JSC),
where the simulations reported in this
paper were primarily performed.
We thank 
P.K. Yeung and Kiran Ravikumar 
for generously providing the data at 
$18432^3$, which was generated
on Summit under the DOE INCITE 2019 award.
AP was supported by the TILT project from ANR
(Contract No. ANR-20-CE30-0035).
\end{acknowledgements}


%
\end{document}